\RequirePackage[T1]{fontenc}
\documentclass[journal]{IEEEtran}
\usepackage{amsmath,amssymb,graphicx,booktabs,url}
\usepackage[hidelinks]{hyperref}
\newcommand{\doi}[1]{\href{https://doi.org/#1}{doi: #1}}
\DeclareMathOperator{\E}{E}
\begin{document}
\title{A Conditional Reference for Sentinel-1 Interferometric Closure Over Wheat Fields}
\author{Shubham K. Singh\thanks{Shubham K. Singh is with Tufts University, Medford, MA, USA.}}
\markboth{Interferometric closure over wheat fields}{Singh: A conditional reference for interferometric closure}
\maketitle
\begin{abstract}
Small-parcel Sentinel-1 interferometric closure reuses a few resolution cells across all three interferograms, confounding physical scattering residuals with finite-sample bias and within-date spatial correlation. We present a finite-support diagnostic framework centered on an exact conditional reference: under blockwise random phase rotations that preserve each block's measured scattering while removing between-block alignment, we derive the closed-form conditional mean $H$ and variance $\E_0|\widehat B - H|^2$ of the ordinary closure product $\widehat B$. We pair this reference with a distinct-block $U$-statistic and footprint-matched, spatially correlated zero-coherence nulls. Across 12 two-texture mixtures at eight looks with estimated powers, $U$ reduced complex-product RMSE in only three mixtures and mean angular error in one. Over Sentinel-1A wheat parcels near Varanasi, India (2023--2024), the median same-block fraction fell by 20.8 percentage points in VV and 11.2 in VH, compared with an 11.0-point decline from footprint geometry alone under independent zero-coherence pixels. Fifteen of 112 parcel--date--polarization tests exceeded nominal block-rotation thresholds, a count reproduced by zero-coherence simulations in 1.5\% of campaigns under moderate spatial correlation and 72.5\% under stronger correlation; zero also outperformed $\widehat B$ and $U$ in complementary half-field prediction in 2024. Footprint-matched spatial diagnostics are therefore essential before attributing small-parcel closure angles to crop growth or soil moisture.

\end{abstract}

\section{Introduction}
A common acquisition phase cancels around a three-date interferometric loop, leaving a nonzero closure residual when ground and canopy scatterers follow distinct phase histories \cite{dezan,yuan}. Because wheat phenology and soil moisture alter radar phase, backscatter, and coherence \cite{yuan,palm,zwieback,ouaadi,singhWCM,singhGPR}, parcel-level closure is an attractive biophysical observable. Interpreting a parcel angle is hazardous, however, when all three interferograms reuse a few spatially correlated resolution cells.

Prior work established closure bias and testing at broader scales: Ansari \emph{et al.} documented systematic phase bias in short-baseline multilooked interferograms \cite{ansari}, Zheng \emph{et al.} linked closure to deformation time-series bias \cite{zheng}, Zwieback \emph{et al.} tested closure against decorrelation noise via an asymptotic Wald statistic \cite{zwiebackWald}, and optical bispectrum methods correct photon and read noise \cite{gordon}. Over smallholder parcels, often only tens of meters wide, asymptotic assumptions break down, and pixel reuse couples finite-sample noise with within-date spatial correlation.

This letter introduces a finite-support diagnostic framework to assess genuine cross-block alignment from sampling and spatial-correlation artifacts: (1) we derive the exact conditional mean $H$ and variance $\E_0|\widehat B - H|^2$ of the ordinary product $\widehat B$ under blockwise random phase rotations; (2) we benchmark a distinct-block $U$-statistic against $\widehat B$ across two-texture mixtures under known and estimated powers; and (3) we pair the conditional reference with footprint-matched, spatially correlated zero-coherence nulls. Evaluated on two Sentinel-1A wheat campaigns near Varanasi, India, the framework demonstrates why footprint-matched diagnostics are required before attributing parcel closure angles to crop or soil changes.

\begin{figure*}[t]
\centering\includegraphics[width=.94\textwidth]{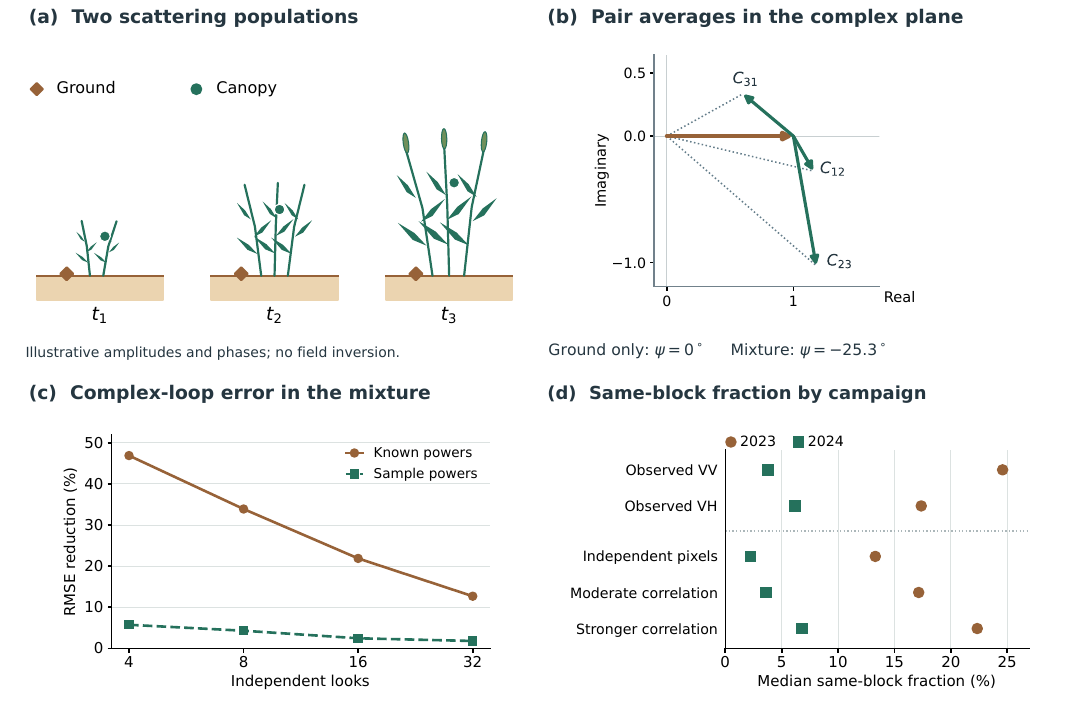}
\caption{Scattering and finite spatial support. (a) Ground and canopy represent independent scattering textures. (b) Each dotted ray points to a pair covariance. A common ground vector (brown) and one pair-specific canopy vector (green) sum to give each endpoint. The coefficients in Section~\ref{sec:loop} generate the plotted angles. (c) Relative reduction in complex-loop RMSE when three distinct looks are used in the mixture in (a),(b). The two series compare normalization by known or estimated powers (40\,000 realizations per look count). (d) Percentage of the mean squared ordinary loop under blockwise random phase rotations accounted for by the squared conditional mean. Circles and squares show the 2023 and 2024 campaign medians. The three reference rows give medians across simulated campaigns on the same footprints with zero temporal coherence and apply to both channels.}
\label{fig:mechanism}
\end{figure*}
\section{From a Scattering Mixture to a Parcel Loop}\label{sec:loop}
Let $z_i$ denote the calibrated complex return at the $i$th acquisition in one polarization. Positive phase follows $C_{ij}=\E[z_i z_j^*]$. In Fig.~\ref{fig:mechanism}(a), take two zero-mean, mutually uncorrelated unit-power textures $s$ and $v$ for the ground and canopy, respectively. Their coefficients $g_i$ and $b_i$ can include a change in amplitude or path length. The resulting covariance and population loop are
\begin{align}
 z_i &= g_i s + b_i v, & C_{ij}&=g_i g_j^*+b_i b_j^*,\label{eq:mix}\\
 B &= C_{12} C_{23} C_{31}, & \psi&=\arg B. \label{eq:B}
\end{align}
A single coherent texture gives $B=|g_1g_2g_3|^2$ whatever the three individual phases. For a numerical example fix all $g_i=1$ and let the canopy amplitudes be $0.4,0.8,1.3$ with phases $0^\circ,60^\circ,140^\circ$. The arguments of $C_{12}, C_{23}, C_{31}$ are then $-13.4^\circ,-40.9^\circ,+29.1^\circ$, yielding $-25.3^\circ$ around the loop. The two polarizations can weight these textures differently. Recovering their contributions from a single angle is underdetermined.

\subsection{The effect of reusing a look}
Use $L$ independent, identically distributed, zero-mean proper complex Gaussian looks with covariance $C$. Write $P_i=C_{ii}$ and let $\widehat C$ be the sample covariance formed with the known zero mean. For the ordinary product $\widehat B=\widehat C_{12}\widehat C_{23}\widehat C_{31}$, expansion of the three sums gives
\begin{align}
 \E\widehat B &= B + S/L + (D+B^*)/L^2, \label{eq:bias}\\
 S &= P_1|C_{23}|^2+P_2|C_{31}|^2+P_3|C_{12}|^2, \nonumber\\
 D &= P_1P_2P_3. \nonumber
\end{align}
To obtain (\ref{eq:bias}), assign a look index to each of the three pair products. Three different indices factor into $B$. When exactly two indices coincide, Gaussian pairing adds one term of $S$ to $B$. When all coincide, the product is $|z_1 z_2 z_3|^2$ and its mean is $D+S+B+B^*$. Counting the assignments gives (\ref{eq:bias}). Even when all population coherences vanish, the expected product is $D/L^2$, positive and real. This is a bias of a complex moment. It supplies no additive correction in degrees.

Spatial dependence can also link distinct pixels when acquisitions are independent. Write $X_{ij}(a)=z_i(a)z_j(a)^*$ and $K_i(a,c)=\E[z_i(a)z_i(c)^*]$. Factoring by acquisition gives
\begin{equation}
\E[X_{12}(a)X_{23}(b)X_{31}(c)]=K_1(a,c)K_2(b,a)K_3(c,b).
\label{eq:spatial}
\end{equation}
For positive real spatial covariances, nearby pixels can contribute a positive term even across a block boundary. Independent rotations of those blocks erase its conditional mean.

\begin{figure*}[t]
\centering\includegraphics[width=.94\textwidth]{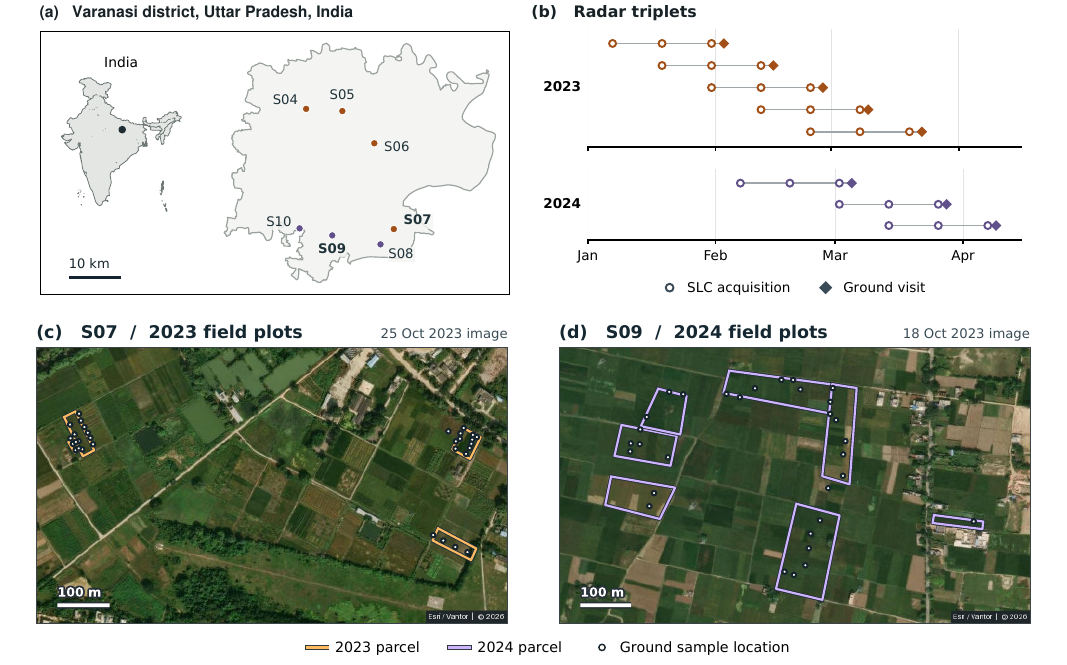}
\caption{Study sites and sampling calendar. (a) Varanasi district and the seven sites used in this study. S04--S07 supplied the 2023 observations and S08--S10 the 2024 observations. (b) Each row shows three 12-day-spaced radar dates followed by a ground visit. (c),(d) Author-mapped wheat parcels and measured locations on Esri World Imagery. The WorldView-3 basemap scenes were captured on 25 October 2023 at S07 and 18 October 2023 at S09. The map grid is approximately 1.78 m at the fields. These dates fall outside the depicted wheat visits. Imagery \copyright~2026 Esri and its licensors. Source: Esri World Imagery, Vantor. North is up.}
\label{fig:map}
\end{figure*}

\subsection{Two treatments of repeated spatial support}
For a field, partition the jointly valid pixels into occupied blocks with weights $w_b=n_b/\sum n_b$. Denote the within-block pair means for $12,23,31$ by $A_b,Q_b,R_b$. The ordinary field product multiplies the three weighted pair means. The version restricted to three distinct blocks is
\begin{align}
 U&=\frac{\sum_{b,c,d\ {\rm distinct}} w_b w_c w_d A_b Q_c R_d}{W_3},\label{eq:u}\\
 W_3&=1-3\sum_b w_b^2+2\sum_b w_b^3. \nonumber
\end{align}
Conditional on fixed weights, independent block estimates with common pair expectations give $\E U=B$ by factorization. This is a weighted form of a third-order $U$-statistic \cite{hoeffding}. Independence is needed for that interpretation. The formula is computable for three or more occupied blocks with valid products. With few or unbalanced blocks $W_3$ can be small and removing the repeated terms can increase variance.

The second treatment asks what the measured field product would look like without alignment \emph{between} blocks. Multiply every pixel at date $i$ in block $b$ by $\exp(\mathrm i\theta_{ib})$, with independent uniform angles by date and block, shared across VV and VH. This preserves all single-date powers, the magnitudes of the three within-block pair means, and each within-block loop. In a term $A_b Q_c R_d$, all rotation factors cancel only for $b=c=d$. Thus
\begin{equation}
 H=\E_0[\widehat B\mid\text{block data}]=\sum_b w_b^3 A_b Q_b R_b. \label{eq:H}
\end{equation}
$H$ can be complex. Subtracting it yields a cross-block contrast with zero conditional mean. The identity
\begin{equation}
 \E_0|\widehat B|^2=|H|^2+\E_0|\widehat B-H|^2 \label{eq:mse}
\end{equation}
quantifies the reference contribution without choosing a coherence threshold. The fraction $|H|^2/\E_0|\widehat B|^2$ is per field and polarization. A one-block field has $\widehat B=H$ for every rotation and contains no between-block information. For direct computation let $a_b=w_b A_b$, $q_b=w_b Q_b$ and $r_b=w_b R_b$. Every index triple except the all-equal ones carries a distinct rotation phase factor. Set $V_a=\sum_b|a_b|^2$ and define $V_q,V_r$ analogously. Orthogonality of these phase factors gives
\begin{equation}
 \E_0|\widehat B-H|^2 = V_a V_q V_r-\sum_b|a_b q_b r_b|^2. \label{eq:var}
\end{equation}
Both moments require only sums over occupied blocks. Their exactness concerns the specified rotations. Physical dependence across blocks can survive in the observations. Terms with two equal indices contribute to $\widehat B-H$ and drop out of $U$.

\section{Wheat Fields, Radar Processing, and Benchmarks}\label{sec:methods}
Thirteen ascending Sentinel-1A relative-orbit-56 acquisitions over Varanasi, India, supply five 2023 and three 2024 triplets with 12/12/24-day spacing (Fig.~\ref{fig:map}). Ground visits followed each triplet's last acquisition by two or three days across disjoint sites (S04--S07 in 2023; S08--S10 in 2024), with neighboring triplets sharing acquisitions.

We processed IW3 single-look complex data in ESA SNAP 10 using precise orbits, complex sigma-nought calibration, terrain-assisted back-geocoding, and enhanced spectral diversity \cite{tops}, registering to the first date. After flat-earth and terrain phase removal, we combined registered amplitudes with phases $0,-\theta_{12},-\theta_{13}$ on a common VV/VH valid-pixel mask. A 10-m inward parcel buffer left 25 and 33 nonempty parcel--dates in 2023 and 2024, respectively, across 11 fields per campaign (eight candidate interiors were empty; 2--51 native pixels in 2023 and 11--380 in 2024).

Pixel centers were grouped into 30-m blocks by integer division of UTM zone 44N coordinates. For Sentinel-1 IW3 (roughly 5~m range by 20~m azimuth), 30~m exceeds one native azimuth resolution cell while preserving multiple occupied blocks inside smallholder parcels (often 30--100~m wide). Finer blocks would increase inter-block correlation, whereas coarser blocks would collapse small 2023 parcels below the three-block minimum for $U$ (or into one-block cases where $\widehat B=H$). Because spatial correlation can still bridge 30-m boundaries, we do not assume strict block independence; instead, we benchmark block-rotation statistics against native-lattice spatially correlated nulls. Dividing each date's returns by its field root mean power yields $T=\hat\gamma_{12}\hat\gamma_{23}\hat\gamma_{31}$, leaving divisors rotation-invariant; (\ref{eq:bias}) and the unbiasedness of $U$ apply prior to power normalization.

Three Monte Carlo suites evaluate estimator error and spatial support: (i) \emph{Gaussian mixtures}: 40\,000 paired draws per setting across 4--32 independent looks (unit powers; coherence magnitudes 0, 0.25, or 0.5; $C_{12},C_{23}>0$, $\arg C_{31}=\pm 45^\circ$), reporting complex RMSE $\sqrt{\E|\widehat B-B|^2}$ and mean absolute shortest-arc error when $B\ne0$; (ii) \emph{block rotations}: 19\,999 rotations per parcel--date for nominal $|\widehat B-H|$ $p$-values and 39\,999 joint rotations sharing draws across reused acquisitions, ground blocks, and channels; and (iii) \emph{zero-coherence footprint nulls}: 16\,384 independent-pixel unit-power and fixed-amplitude phase-randomized draws for same-block fractions, 8192 native-lattice campaigns per spatial model (median parcel RMS mapping displacement 2.85~m) for pair-coherence medians, and 4096 full campaigns per separable azimuth/range correlation model $\rho$ with 2000 inner rotations (verified at 19\,999 rotations on 512 campaigns) for independent ($\rho_p=0$) and correlated ($\rho_p=0.70$) polarizations.

\section{Complex-Product Error and Spatial Support}
\subsection{Bias--variance trade-off in a known scattering mixture}
Drawing independent Gaussian ground and canopy textures in (\ref{eq:mix}) gives the normalized population loop $0.22661-0.10724\mathrm i$. At eight looks with known powers, $\widehat B$ has expected real part $0.42743$; using distinct looks via $U$ removes this displacement and cuts complex RMSE by 33.9\% (0.6462 to 0.4272). With estimated powers, however, complex RMSE falls by only 4.3\% (0.2043 to 0.1956) while mean absolute angular error rises from $21.1^\circ$ to $40.7^\circ$: a point can move closer to the true complex loop while pointing farther from its direction.

Varying the three canopy amplitudes together (multipliers 0.5, 1, or 2) and their phases together (0, 0.5, 1, or 1.5) at eight looks with estimated powers, $U$ reduced complex RMSE in only three of 12 mixtures and angular error in one. The gain in Fig.~\ref{fig:mechanism}(c) is mixture-dependent and does not guarantee improved angle retrieval.

\subsection{Footprint geometry and same-block dominance}
Median 30-m block occupancy rose from three in 2023 to ten in 2024. Two of 25 nonempty 2023 cores occupied one block (F23 on 31 January; F32 on 8 March), leaving 23 parcel--dates for alignment testing and 20 for $U$, whereas all 33 in 2024 supported both. F32's three pixels gave $+54.3^\circ$ in VV and $-51.7^\circ$ in VH, both invariant under block rotation because $\widehat B=H$.

Under zero temporal coherence, finite samples still yield positive coherence magnitudes ($|\hat\gamma|^2\sim\mathrm{Beta}(1,N-1)$ for $N$ independent pixels \cite{touzi}). Table~\ref{tab:support}(a) compares measured 12-, 12-, and 24-day pair medians against footprint-matched zero-coherence nulls. Independent pixels alone gave 0.215 (2023) and 0.088 (2024). Under stronger spatial correlation (0.50/0.75), the 2023 null median (0.356) lies within the measured range (0.339--0.377), while the weakest 2024 pair (24-day: 0.114 VV, 0.109 VH) matches the moderate-correlation null (0.115). Because temporal coherence would raise rather than lower these magnitudes, the 0.50/0.75 and 0.65/0.65 nulls (0.171--0.180) overstate spatial dependence for this pair; for 2024 the moderate model is the better-supported reference.

\begin{table}[t]\centering
\caption{Measured pair coherence and geometry-matched simulations with zero temporal coherence. The four reference columns in (a) correspond, in order, to the spatial models in (b).}
\label{tab:support}\small\setlength{\tabcolsep}{2.5pt}
\begin{tabular}{@{}llrrrrrr@{}}\toprule
\multicolumn{8}{c}{(a) Campaign median $|\hat\gamma|$ by acquisition pair}\\
& & \multicolumn{2}{c}{Measured} & \multicolumn{4}{c}{Footprint-matched null median}\\
\cmidrule(lr){3-4}\cmidrule(l){5-8}
Year ($n$) & Pair & VV & VH & IID & Mod. & Strong & 0.65/0.65\\\midrule
2023 (23) & 12 & 0.347 & 0.377 & 0.215 & 0.263 & 0.356 & 0.331\\
 & 23 & 0.339 & 0.358 & 0.215 & 0.263 & 0.356 & 0.331\\
 & 13 & 0.347 & 0.373 & 0.216 & 0.262 & 0.356 & 0.332\\
\addlinespace[2pt]
2024 (33) & 12 & 0.188 & 0.167 & 0.088 & 0.114 & 0.180 & 0.171\\
 & 23 & 0.148 & 0.181 & 0.088 & 0.115 & 0.180 & 0.172\\
 & 13 & 0.114 & 0.109 & 0.088 & 0.115 & 0.180 & 0.171\\
\bottomrule\end{tabular}\par\smallskip\setlength{\tabcolsep}{4pt}
\begin{tabular}{lrrr}\toprule
\multicolumn{4}{c}{(b) Zero temporal coherence, 112 tests per campaign}\\
& \multicolumn{1}{c}{Mean count} & \multicolumn{2}{c}{Campaigns with count $\geq15$}\\
\cmidrule(lr){3-4} Az. / range $\rho$ & $\rho_p=0$ & $\rho_p=0$ & $\rho_p=0.70$\\\midrule
0 / 0 & 5.63 & 0.07\% & 0.12\%\\
0.25 / 0.50 & 7.96 & 1.49\% & 1.95\%\\
0.50 / 0.75 & 16.90 & 72.53\% & 70.68\%\\
0.65 / 0.65 & 18.49 & 84.96\% & 83.42\%\\
\bottomrule\end{tabular}
\end{table}

Figure~\ref{fig:mechanism}(d) isolates footprint geometry in the same-block fraction $|H|^2/\E_0|\widehat B|^2$. Its median fell from 24.58\% to 3.81\% in VV ($-20.8$ percentage points) and from 17.37\% to 6.17\% in VH ($-11.2$ points). On identical masks, independent zero-coherence pixels dropped from 13.30\% to 2.25\%, an 11.0-point decline from geometry alone (fixing measured amplitudes with random phases gave 15.65\% to 2.75\% in VV and 14.66\% to 2.68\% in VH). On 20- and 40-m grids the contrast persisted: the 2023 VV median ranged 19.9--24.9\% across the three grids, versus 3.0--7.9\% in 2024. Adding separable azimuth/range correlation raised simulated medians to 17.15\%/3.61\% (0.25/0.50) and 22.34\%/6.79\% (0.50/0.75). In 2024, seven of 33 observations per channel exceeded the independent-pixel 97.5th percentile, versus one under stronger correlation.

Matching standardized master-date amplitude correlations highlights the identifiability limit. In 2024, measured adjacent-pixel azimuth/range amplitude correlations were 0.374/0.363 (VV) and 0.357/0.344 (VH), versus 0.202/0.505 for the 0.50/0.75 model. Complex lag-one correlations of 0.65/0.65 matched 2024 amplitudes (0.370/0.365) but overstated 2023 VV azimuth dependence (0.322/0.273 vs. 0.120/0.251 VV and 0.248/0.296 VH measured). Because two amplitude moments leave the complex spatial covariance underdetermined, and shared fields and acquisitions couple campaigns, year-specific attribution remains constrained.

\subsection{Alignment exceedances without temporal coherence}
For $|\widehat B-H|$, 15 of 112 parcel--date--polarization tests had nominal $p<0.05$, though none passed Holm adjustment \cite{holm}. Under joint block rotations preserving shared acquisitions and blocks, the mean count was 5.60 (95\% range 2--11), with 30 of 39\,999 draws reaching $\ge 15$ ($p=0.000775$). Even a single coherent texture with $\arg B=0$ departs from block rotations, and the statistic tests neither $\arg B$ nor individual pair coherences.

Crucially, native-lattice zero-coherence simulations can reproduce this count via (\ref{eq:spatial}). In Table~\ref{tab:support}(b) ($\rho_p=0$), 1.49\% of moderate-correlation campaigns (61/4096) and 72.53\% of strong-correlation campaigns (2971/4096; median 17, 95\% range 10--25) reached $\ge 15$ exceedances, confirmed by full 19\,999-rotation checks (13 and 374 of 512 campaigns). The 0.65/0.65 model reached $\ge 15$ in 84.96\% of campaigns, and correlating polarizations ($\rho_p=0.70$) barely altered these rates. Thus, zero-coherence returns can reproduce the observed count under stronger spatial correlation, but rarely (1.5\%) under the moderate correlation favored by the 2024 24-day pair; separating sensor or processing correlation from ground scattering requires external information.

\subsection{Internal spatial predictability}
To test field-scale spatial consistency directly, we split cores with at least six blocks into contiguous halves along their first principal axis. Each half's $\widehat B$ or $U$ predicted the complementary half's $U$, normalizing complex differences by the predictor half's power product and averaging squared errors symmetrically. Across 26 eligible 2024 parcel--dates (nine fields; only three 2023 cores qualified), zero beat both estimators: RMSE for zero, $\widehat B$, and $U$ was 0.0310, 0.0330, and 0.0379 in VV, and 0.0362, 0.0394, and 0.0403 in VH. Weak loops, finite sampling, and within-field heterogeneity jointly limit spatial transferability.

Comparing normalized core alignment against angular agreement between each core and its disjoint 10-m edge strip across all 33 parcel--dates in 2024 likewise gave negligible Spearman correlations: $0.044$ in VV (95\% field-clustered bootstrap interval $-0.34$ to $0.39$) and $-0.135$ in VH ($-0.38$ to $0.10$), conditioned on the sampled sites and acquisitions.

\section{Diagnostic Protocol and Conclusion}
Before attributing a parcel closure angle to crop growth or soil moisture, we recommend a four-step finite-support protocol:
\begin{enumerate}
\item \emph{Audit spatial support:} Record the common pixel mask and block occupancy at or above native resolution; flag one-block ($\widehat B=H$) and $<3$-block ($U$ undefined) parcels.
\item \emph{Compute the conditional reference:} Report $H$, same-block fraction $|H|^2/\E_0|\widehat B|^2$, and cross-block contrast $|\widehat B-H|$ via (\ref{eq:H})--(\ref{eq:var}) to gauge sample-reuse dominance.
\item \emph{Benchmark against spatial nulls:} Compare pair-coherence medians and alignment counts against footprint-matched zero-coherence simulations across spatial correlations compatible with measured amplitude lags and with the lowest pair-coherence median.
\item \emph{Check internal consistency:} On parcels with $\ge 6$ blocks, verify whether half-field loops beat zero in predicting complementary halves.
\end{enumerate}

In Varanasi Sentinel-1A wheat data, 15 alignment exceedances were unusual under block rotations yet common under strongly correlated zero-coherence nulls permitted by (\ref{eq:spatial}) and rare under the moderate model favored by the 2024 24-day pair, and zero won 2024 half-field prediction in both channels. Whenever closure significance depends on the spatial correlation model, physical interpretation remains unresolved, and no universal pixel-count threshold guarantees angle reliability.

\end{document}